\documentclass[a4paper,aps,prl,twocolumn]{revtex4}

\usepackage{graphicx} 
\usepackage[usenames,dvipsnames]{xcolor}
\usepackage{tikz}
\usepackage{amsmath}
\usepackage{amssymb}
\usepackage{natbib}
\usepackage{color}
\usepackage{psfrag}
\usepackage{wasysym}
\usepackage{float}
\usepackage{epstopdf}
\usepackage[normalem]{ulem}
\usepackage{pgfplots}
\usepackage{pgfplots, pgfplotstable}
\usepackage{textcomp}
\usepackage{hyperref}
\usepackage{fourier}
\usepackage{arcs}
\usetikzlibrary{pgfplots.groupplots}
\usepgfplotslibrary{fillbetween}
\usetikzlibrary{shapes.arrows}

\definecolor{maroon}{RGB}{128,0,0}

\begin{document}

\title{Multistable Kirigami for Tunable Architected Materials}
\author{Yi Yang$^1$, Marcelo A. Dias$^2$, and Douglas P. Holmes$^1$}
\affiliation{\footnotesize$^1$ Mechanical Engineering, Boston University, Boston, MA, 02215, USA\\ \footnotesize$^2$ Department of Engineering, Aarhus University, 8000 Aarhus C, Denmark}

\date{\today}

\begin{abstract}
In nature, materials such as ferroelastics and multiferroics can switch their microstructure in response to external stimuli, and this reconfiguration causes a simultaneous modulation of their material properties. Rapid prototyping technologies have enabled origami and kirigami-inspired architected materials to provide a means for designing shape-shifting structures, and here we show how multistable structures inspired by kirigami provide novel design criteria for preparing mechanical metamaterials with tunable properties. By changing the geometry of kirigami unit cells, we obtain multistable kirigami lattice structures endowed with a bistable snap-through mechanism. We demonstrate the precise control of material stiffness, along with the ability to tune this property {\em in situ} by locally and reversibly switching the unit cell configurations. We anticipate these mechanical metamaterials will provide a platform to achieve {\em in situ} tunable electrical, optical, and mechanical properties for a variety of applications in multifunctional materials, two-dimensional materials, and soft robotics.
\end{abstract}


\maketitle

Materials with reconfigurable architecture may exhibit tunable electrical, optical, and mechanical properties. Notable examples are found in ferroelasticity\cite{Salje2012} and multiferroics\cite{Cheong2007}. Applying an external stimulus, such as a stress/strain field, magnetic field, or electric field results in a structural or electronic phase transformation in the atomic scale which modulates the bulk material properties. Recent investigations on reconfigurable and programmable architected materials provide a new opportunity to attain tunable material properties by systematically programming the microstructure of a constituent material\cite{Overvelde2017,Rocklin2017,Coulais2016,Babak2016,Florijn2014,Silverberg2014}. The mechanical properties of these architected materials depend on the topology and geometry of the substructure, and are typically independent of the constituent's chemical composition. By introducing controllable morphological structures into the unit cell, reprogrammable and reconfigurable metamaterials can be achieved. Among various types of architected materials, kirigami and origami inspired metamaterials attracted tremendous attention due to their robust and straightforward ability to transform 2D sheets into 3D structures\cite{Shyu2015,YTang2017,Cho2014,Blees2015,Florijn2014, MKo2016, AaronL2015, Zhang2015, YTang2015, Faber2018, Overvelde2017, Boatti2017, Filipov2015, Silverberg2014, Chen2016, Hwang2018}. However, compared with origami-inspired metamaterials, which have been extensively studied\cite{Faber2018, Overvelde2017, Boatti2017, Filipov2015, Silverberg2014, Florijn2014}, understanding the behavior of kirigami structures is limited\cite{Shyu2015, Ahmadkatia2017, Cho2014, MKo2016, Sussman2015,moshe2018nonlinear}. Hence, there remains a significant opportunity to advance the design of kirigami-based metamaterials with tunable material properties.

In this study, through a combination of experiments, finite element (FEA) simulations, and theoretical analyses, we demonstrate how a multistable microstructure inspired by kirigami provides a novel approach to designing mechanical metamaterials with tunable material properties. By changing the spacing between the adjacent slits in the conventional linear parallel cutting patterns, we obtain multistable kirigami lattice structures composed of repeating unit cells whose structure is energetically metastable with two local minima, thereby endowing it with a snap--through mechanism. Each local stable state is associated with a corresponding structural configuration. By applying an external perturbation, we can manipulate the unit to switch between the two stable configurations rapidly and reversibly. Furthermore, we showcase an {\em in situ} tunability of material stiffness using this multistable kirigami. We also demonstrate that the multistable kirigami mechanism is material independent and scale invariant, enabling it to be integrated with stimuli-responsive materials, 3D printing techniques, or combined with origami to be applied in multifunctional materials, two-dimensional materials, soft robotics, and biomedicine.

\begin{figure*}
\centering
\includegraphics[width=1\textwidth]{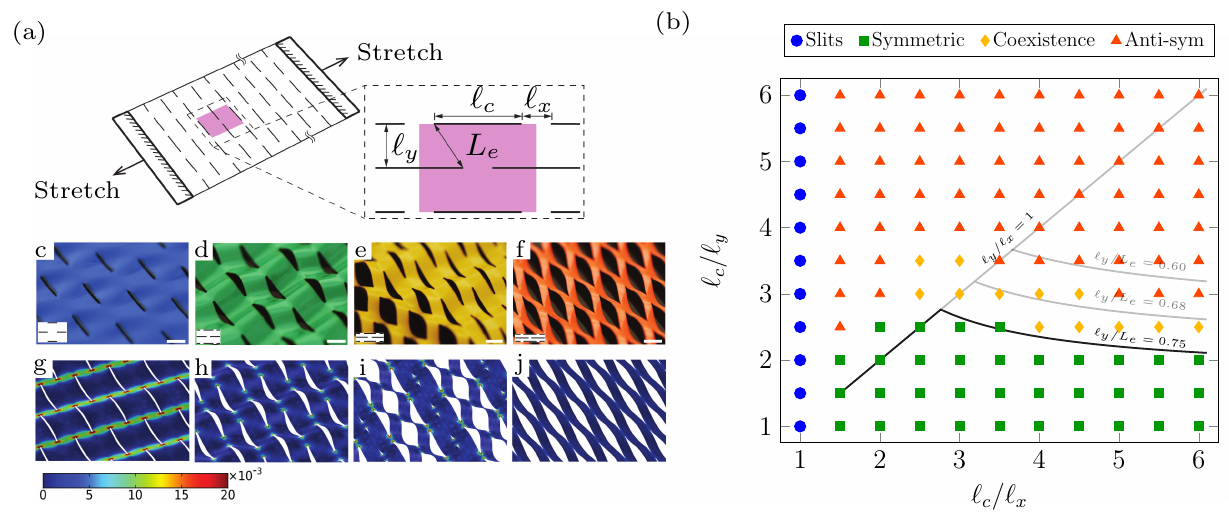} 
\caption{(a) Schematic diagram for a typical kirigami specimen under uni-axial stretching with the defined geometric parameters (unit cell highlighted in pink). (b) Phase diagram obtained from experiments demonstrating the variation in buckling configurations as the geometric parameters $\ell_c/\ell_x$ and $\ell_c/\ell_y$ are varied. Experimental images (c-f) and the corresponding FEA simulations (g-j) illustrating typical postbuckled 3D lattice structures. (c,g) A set of single slit in an array ($\ell_c/\ell_x=\ell_c/\ell_y=1.0$). (d,h) symmetric  configuration ($\ell_c/\ell_x=3.0$, $\ell_c/\ell_y=2.0$). (e,i) Coexistence of the symmetric and anti-symmetric configuration ($\ell_c/\ell_x=3.0$, $\ell_c/\ell_y=3.0$). (f,j) Anti-symmetric configuration ($\ell_c/\ell_x=5.0$, $\ell_c/\ell_y=5.0$). The scale bars in (c) to (f) represent $5$ mm. Color bar represent the magnitude of von Mises stress normalized by the material's elastic modulus.}
\label{fig1}
\end{figure*}

To demonstrate this concept, we utilized polyethylene terephthalate sheets (Dupont Teijin Film, McMaster--Carr) with Young's modulus $E = 3.5$ GPa, Poisson's ratio $\nu = 0.38$ and thickness $h = 0.127$ mm. Kirigami patterns can be represented by their unit cell. The unit cell geometry of the linear parallel cut\cite{Shyu2015} is characterized by the cut or slit length ($\ell_c$), and the spacing between two adjacent slits along the transverse direction ($\ell_x$) and longitudinal direction ($\ell_y$), as demonstrated in Fig.~\ref{fig1}a. These parameters can be combined and non-dimensionalized as $\ell_c/\ell_x$ and $\ell_c/\ell_y$ \cite{Shyu2015} to describe the two-dimensional geometry of the unit cell. For the sake of simplicity, we used a constant slit length and varied $\ell_x$ and $\ell_y$ to achieve a wide variation of the unit cell geometry. Note that every unit cell is symmetric about its longitudinal and transverse center line. The overall geometry of a test specimen was selected based on the ASTM D882 standard with a constant slit length of 15.2 mm.  

We performed a systematic set of experiments by varying $\ell_c/\ell_x$ and $\ell_c/\ell_y$, both from $1.0$ to $6.0$ with an increment of $0.5$ to investigate the structural configuration of the kirigami metamaterials via uniaxial tensile tests. As we quasi--statically increase the magnitude of stretch ($0.1$ mm/s), the deformation of the material continuously transitions from an in-plane deformation to an out-of-plane deformation at a critical point for all the specimens tested. As reported in the literature \cite{Ahmadkatia2017, MKo2016, Holmes2017}, we identify the transition from the in-plane deformation to the out-of-plane deformation as buckling. The load triggering this transition is the critical buckling force and the corresponding deformation is referred to as the buckling configuration(Fig.S1\cite{SM2018}). In Fig.~\ref{fig1}b, we construct a quantitative phase diagram via experiments to demonstrate the 3D buckling configuration as a result of varying the 2D cutting patterns.

When $\ell_c/\ell_x=1$, as shown in Fig.~\ref{fig1}c, there is no material overlap between two adjacent slits along the longitudinal direction which leads to a weak interaction between slits in the array. In this case, individual slits create a geometric incompatibility, which results in an out-of-plane buckling of the sheet around the slit. Therefore, the buckling configuration is similar to that of the buckled sheet with one single slit\cite{Holmes2017}. As we enhance the slit interactions by increasing $\ell_c/\ell_x$ and $\ell_c/\ell_y$, the buckling configurations transitions from a symmetric configuration (Fig.~\ref{fig1}d) to an anti-symmetric configuration (Fig.~\ref{fig1}f) through a configuration coexistence (Fig.~\ref{fig1}e). When the deformed shape of the unit cell is symmetric or anti-symmetric about its transverse axis, it is referred to as {\em symmetric configuration} or {\em anti-symmetric configuration}, respectively. The anti-symmetric configuration is commonly observed and has been reported in the literature\cite{MKo2016,Shyu2015, Blees2015, AaronL2015}, while, here we systematically investigate the symmetric configuration for the first time. Fig.~\ref{fig1}g--j show the stress distribution for the four typical deformation configurations using FEA simulation\cite{SM2018}. As expected, we observed stress concentration at the tips of the slit and relative lower stress elsewhere. These spots would be crucial when designing the kirigami metamaterials. 

\begin{figure*}
\centering
\includegraphics[width=0.95\textwidth]{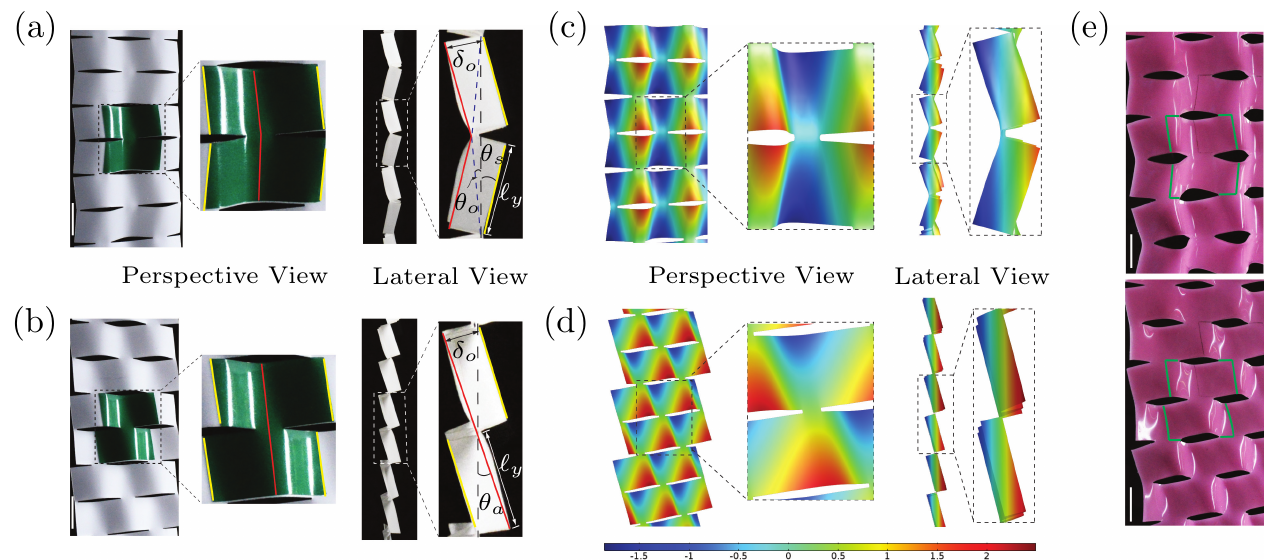}
\caption{(a) and (b), Experimental images demonstrating the reversible configuration transformation between symmetric (a) and anti-symmetric (b) of the bistable unit cell. Unit cell is highlighted in green with its undeformed and deformed longitudinal center line marked in black and red, respectively. (c) and (d), FEA simulations demonstrating the out-of-plane deformation for the symmetric (c) and anti-symmetric (d) of the bistable unit cell. Unit of the color bar is in mm. (e) Experimental images demonstrating sample fabricated using Vinylpolysiloxane (Elite Double 8, Zhermack, a silicone-based polymer, Young's modulus $\approx 0.2$ Gpa, unit cell in green). Scale bars represents $10$ mm.}
\label{fig2}
\end{figure*}

What is significant about the symmetric lattices is that each unit cell is bistable, {\em i.e.} if the post-buckling configuration of a unit cell is symmetric, it can reversibly switch between symmetric and anti-symmetric configurations. In contrast, if the anti-symmetric configuration forms following buckling, it is monostable (Fig.S2\cite{SM2018}). Bistability appears when there are two distinct local energetic minima. If these minima are local but unequal in magnitude, the system is typically referred to as {\em metastable}. To switch between these two distinct configurations as demonstrated in Fig.~\ref{fig2}, the transverse axis of symmetry of a unit cell acts as an elastic hinge allowing the bottom segment to rotate, which gives rise to a limit-point instability. We can trigger each bistable unit cell to switch reversibly by external perturbation such as mechanical indentation\cite{SM2018}. As shown in Fig.~\ref{fig2}, when the elastic hinges are undeformed (anti-symmetric mode), they all stay in the same plane during the stretching process. Therefore, a tilting angle $\theta_a$ will be used to describe the kinematics. However, when the elastic hinges rotate (symmetric mode), the elastic hinges are alternatively distributed on two adjacent parallel planes. Hence, two tilting angles $\theta_o$ and $\theta_s$ will be used to describe the kinematics. These observations are also quantitatively validated using FEA simulation. In addition, we show that this bistability is dictated by geometry and independent of material. As shown in Fig.~\ref{fig2}e, with the same geometric parameters we reproduce the same 3D structure using polyvinylsiloxane whose elastic modulus is $10^4$ times less than that of the polyethylene terephthalate. 

\begin{figure}
\centering
\includegraphics[width=0.475\textwidth]{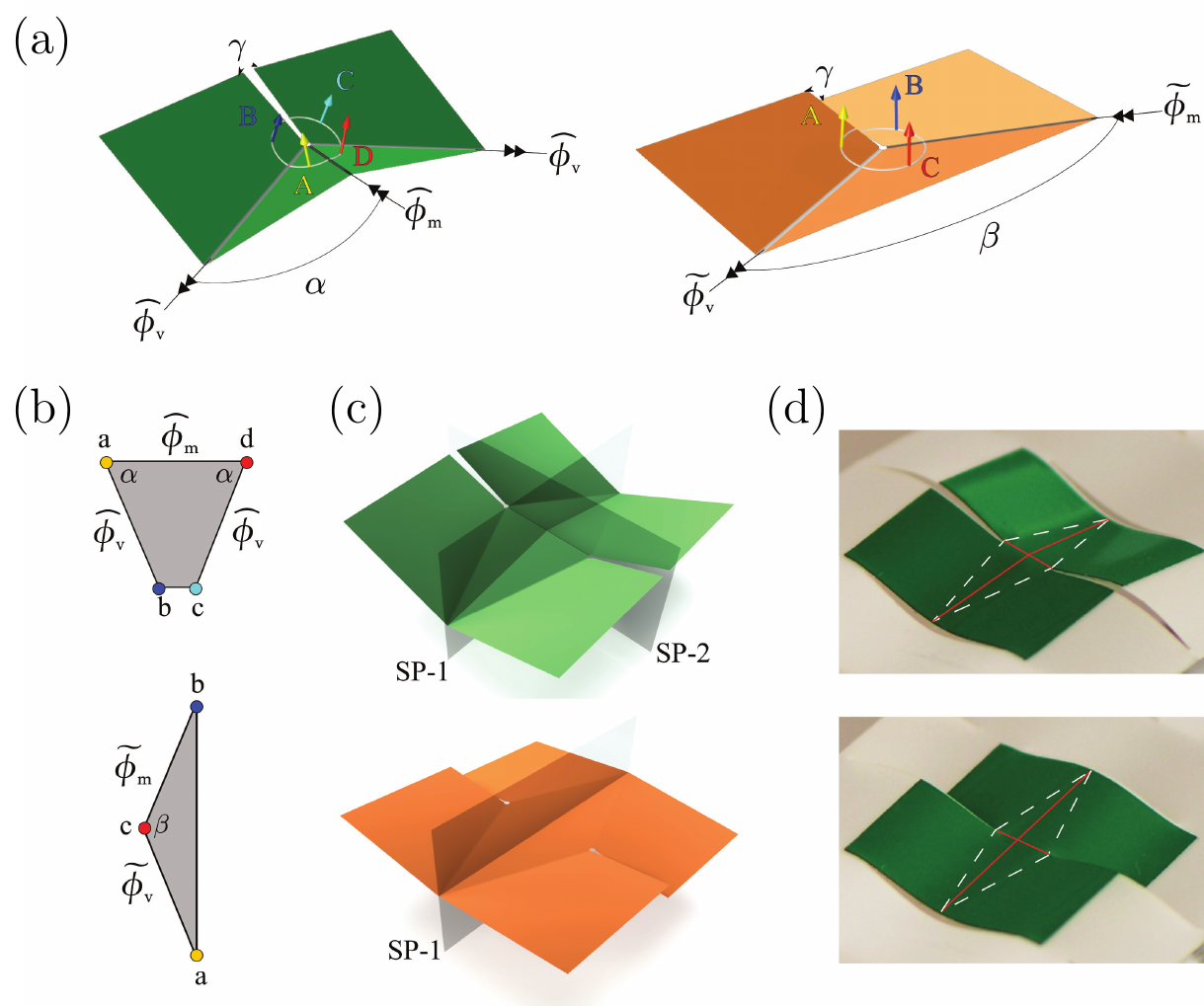}
\caption{These images show the symmetric and anti-symmetric unit cells, represented in terms of sharp folds. (a) Each unit cell is broken down into the \emph{e}-cone motif; for the same angle $\gamma$, two states are adopted. The geometry is given by mountain folding angles $\widearc{\phi}_{\mbox{\tiny m}}$ ($\widetilde{\phi}_{\mbox{\tiny m}}$), valley folding angles $\widearc{\phi}_{\mbox{\tiny v}}$ ($\widetilde{\phi}_{\mbox{\tiny v}}$), and angles between creases $\alpha$ ($\beta$). A path around the cut tip is drawn and normals to each flat face are shown {\bf(A, B, C, D)}. (b) The spherical images of the Gauss map trace polygons on the unit sphere for each configuration. In this spherical image, the normals are mapped onto the polygons vertices, while edge sizes are equal to the folding angles. (c) The symmetric unit cell has more symmetries than anti-symmetric, which is seen from symmetry planes SP-1 and SP-2. (d) Experimental images showing symmetries in the unit cells with smooth bending.}
\label{fig3}
\end{figure}

Next, we rationalize the existence of the bistable mechanism by considering the modes of deformation of a single cut in a sheet, which are described by their fundamental building block: the \emph{e}-cone motif (\emph{excess} angle cone)~\cite{Muller2008}. The unit cells shown in Fig.~\ref{fig2}a, either in the symmetric or anti-symmetric cases, can be seen as the interaction of two neighboring \emph{e}-cones. To better understand the stability of these unit cells, we utilize an approximate analysis that replaces the smooth bending deformations with sharp folds, thus creating mechanisms suitable for the analysis provided by the Gauss map~\cite{Farmer2005,Seffen2016}. Given a fixed opening angle $\gamma$, a single \emph{e}-cone may adopt two possible configurations, the symmetric and anti-symmetric building blocks of each unit cell. As shown in Fig.~\ref{fig3}a--b, the Gauss map takes the normals along a path enclosing the source of gaussian curvature, the tip of the cut, and maps them to the unit sphere. For the sake of simplicity, we consider small enough folding angles so that the spherical images can be approximated to the flat geometry~\cite{Farmer2005}, as depicted in Fig.~\ref{fig3}b. It is known that the area enclosed in the spherical image is proportional to the gaussian curvature, and that this area is also numerically equal to the opening angle $\gamma$, which can be calculated from simple planar geometry given the parameters: a mountain folding angle $\widearc{\phi}_{\mbox{\tiny m}}$ ($\widetilde{\phi}_{\mbox{\tiny m}}$), valley folding angles $\widearc{\phi}_{\mbox{\tiny v}}$ ($\widetilde{\phi}_{\mbox{\tiny v}}$), and the angle between them $\alpha$ ($\beta$). Here we use the $\widearc{(\cdot)}$ and $\widetilde{(\cdot)}$ notation to distinguish between variables within symmetric and anti-symmetric configurations, respectively. Since both symmetric and anti-symmetric configurations can be realized with the same amount of gaussian curvature (by ensuring that the spherical images enclose the same area), the energetic difference when comparing these two mechanisms must be at the level of bending energy. For these discrete systems, bending energies are accounted for by the energies stored in the folds, and they are proportional to the folding angles and the length of the folds. In Fig.~\ref{fig3}c, the symmetric unit cell is shown, including two symmetry planes SP-1 and SP-2, while the anti-symmetric unit has a reduced symmetry given only by the plane SP-1. The minimal number of folds allowed by each state differs by one; the symmetric unit cell having four valleys and one mountain and the anti-symmetric unit cell having two valleys and two mountains. Hence, the bistability arises from the additional plane that the symmetric unit cell can bend about. This is the two dimensional analogue to the buckling of an Euler strut with one hinge versus an Euler strut with two hinges~\cite{Thompson1973,Thompson1984}, where the former has one stable post-buckling shape, while the latter has two. In this classical problem, the minimum energetic state dictates that the two--hinge strut will buckle into a mode shape that is symmetric about the strut's center~\cite{Thompson1973, Thompson1984}, which is consistent to what we observe in our experiments. 

With this qualitative understanding of the origin of the unit cell bistability, we now aim to determine how the geometric parameters $\ell_c/\ell_x$ and $\ell_c/\ell_y$ dictate the postbuckling configuration by evaluating the deformation kinematics and strain energy in each configuration. For this analysis, we return to the smooth bending deformations observed experimentally. Projecting from the side (Fig.~\ref{fig2}), the out-of-plane deflection $\delta_o$, the undeformed plane (black dashed line) and the longitudinal center line (red solid line) forms a right triangle. The kinematic relation is calculated as $\delta_o = \ell_y \tan (\theta_s+\theta_o)$ and $\delta_o = \ell_y \tan (\theta_a)$ for the symmetric and anti-symmetric configuration, respectively. The symmetric configuration can be viewed as four parallelogram plates under bending and two elastic hinges under rotation. By using the beam approximation for skew plate, we transfer the parallelogram plate (skew plate) into a beam with the width as $\ell_y$ and the effective length as\cite{SM2018}
\begin{equation}
\label{Le}
L_e = \sqrt{\left(\frac{\ell_c-\ell_x}{2}\right)^2+\ell_y^2}
\end{equation} 
Since the slope of the beam in the SP-2 plane (Fig.3c) does not change at the boundaries (Videos\cite{SM2018}), the beam is approximate as being clamped on both ends. When the stretching energy and torsion energy induced by the bending of the parallelogram plate is negligible comparing with the out-of-plane bending energy, the total potential energy in the post-buckled unit cell can be expressed as $\Pi = U_b+U_h-W$, where, $U_b$ is the total bending energy of the parallelogram plates, $U_h$ is the total elastic hinge energy and $W$ is the work done by the external tensile force. The configuration selection is determined by the competition between the bending energy and the hinge energy for given geometric parameters. To quantitatively interpret the buckling configuration selection, we utilize the energy ratio between the bending energy and the hinge energy as\cite{SM2018}
\begin{equation}\label{RU}
\frac{U_b}{U_h} \sim \frac{\ell_y}{\ell_x}\frac{\ell_y}{L_e}
\end{equation}
In general, if the total bending energy consumed by out-of-plane bending is significantly larger than the hinge energy, anti-symmetric will be selected, as it is at the lower energy state. In Eq.~\ref{RU}, the energy ratio depends on two geometric parameters, $\ell_y/\ell_x$ and $\ell_y/L_e$. Since $\ell_y/L_e=\ell_y/\sqrt{(\frac{\ell_c-\ell_x}{2})^2+\ell_y^2}$ will always be less than $1$, to ensure $U_b$ is greater than $U_h$, $\ell_y/\ell_x$ has to be greater than $1$. With the assistance of the phase diagram (Fig.~\ref{fig1}b), we report that when the geometric parameter satisfies both $\ell_y/\ell_x<1$ and $\ell_y/L_e>0.75$, the bistable symmetric lattice is obtained.

\begin{figure*}
\centering
\includegraphics[width=1\textwidth]{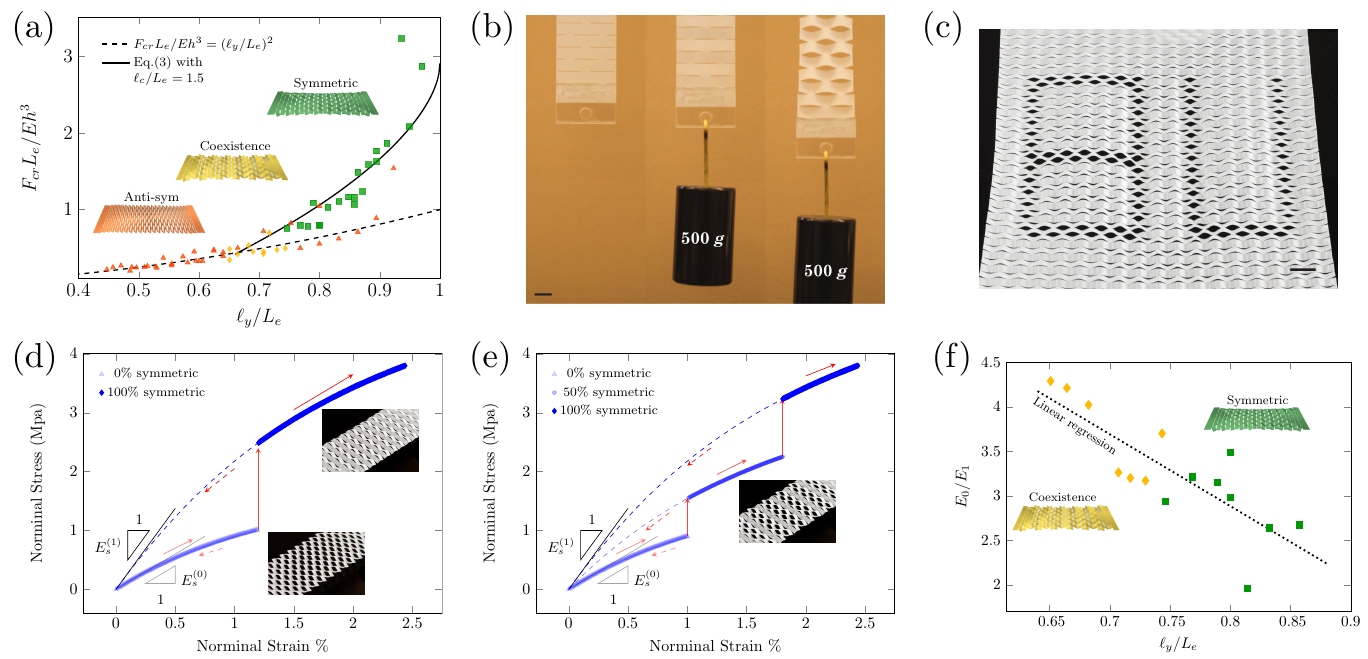}
\caption{(a) Non-dimensionalized critical buckling force varying with $\ell_y/L_e$. (b) Images demonstrate tuning stiffness to lift $500$ g weight using bistable kirigami unit cell. (c) Image showing  selectively tuning unit cells to form capital letters ``B\ U''. Stress-strain curves demonstrating \textit{in situ} tunability of stiffness through a one step tuning (d) and two steps tuning (e) for a specimen with $\ell_c/\ell_x=3.0$ and $\ell_c/\ell_y=2.5$. Red solid line shows the loading path and red dashed line shows the unloading path. (f) Stiffness ratio varying with $\ell_y/L_e$. The slope of the solid line is $-8.065$ and $R^2=65.22\%$ for the linear regression. Scale bars in (d) and (e) represent $10$ mm.}
\label{fig4}
\end{figure*}

Since a pre-stretching load needs to be applied to produce the 3D lattice, to guide the rational design, we calculate the critical buckling force ($F_{cr}$) by minimizing the total potential energy ($\Pi$) with respect to the tilting angles($\theta_o$ and $\theta_s$). The non-dimensionalized critical buckling force for the symmetric configuration is expressed as 
\begin{equation}\label{Fs}
\frac{F_{cr}L_e}{Eh^3}=\frac{1}{2}\left[\frac{\ell_y^2}{L_e^2}+\frac{\ell_x}{L_e(1-\nu)}+\frac{\ell_y}{L_e}\sqrt{\frac{\ell_y^2}{L_e^2}+\frac{2\ell_x}{L_e(1-\nu)}}\right]
\end{equation}
Note that when $\ell_x/L_e$ is negligible, Eq.~\ref{Fs} yields to $F_{cr}L_e/Eh^3=(\ell_y/{L_e})^2$, the non-dimensionalized critical buckling force for the anti-symmetric configuration. In Fig.~\ref{fig4}a, we plot the non-dimensionalized critical buckling force as a function of $\ell_y/L_e$ for specimens with the geometrical parameters in a range of $2.5<\ell_c/\ell_x<5.0$ and $1.5<\ell_c/\ell_y<5.0$, showing our theoretical prediction agrees very well with the experimental data. Here, since Eq.~\ref{Fs} is a function of both $\ell_x/L_e$ and $\ell_y/L_e$, we impose a geometry constraint, $\ell_c/L_e=1.5$, to express $\ell_x/L_e$ in terms of $\ell_y/L_e$ \cite{SM2018}. 

From the perspective of strain energy, both the symmetric and the coexistence configuration are in a state of metastability. The external energy required to cross the energy barrier between the two distinct configurations are not equal and controlled by the unit cell geometry. This ability to tune the energy barrier between the two stable states by geometrical parameters provides an opportunity in designing rapid reconfigurable structures and architected materials. As a result, the reconfiguration of material substructure may lead to modulation of its properties, such as stiffness, Possion's ratio, refraction, and transmittance.

Next, we demonstrate that by assembling the individual bistable unit cell into multistable mechanical metamaterials, we can develop a new strategy to achieve {\em in situ} tunability of the material's stiffness. To generate a multistable lattice, we prescribe a load which is two times of the corresponding critical buckling force. This pre-stretch is held constant in the experiments when measuring the effective stiffness of the resulting structures. As shown in Fig.~\ref{fig4}b, transforming $100\%$ of the unit cells from the symmetric configuration to the $0\%$ symmetric (equivalent to anti-symmetric) configuration significantly reduced the stiffness. The stiffness corresponding to the $100\%$ symmetric and $0\%$ symmetric configuration are the tuning upper and lower bounds, respectively (Fig.S3\cite{SM2018}). We can switch any individual unit cell between its symmetric and anti-symmetric configuration homogeneously or nonhomogeneously to form the desired structure, such as two capital letters ``B\ U'' in Fig.~\ref{fig4}c. Fig.~\ref{fig4}d and Fig.~\ref{fig4}e demonstrate {\em in situ} tuning the stiffness of the multistable kirigami specimen from $0\%$ symmetric to $100\%$ symmetric directly, and through a step of $50\%$ symmetric (homogeneous along the longitudinal direction), respectively. We also note that the strain-stress curve shows hysteresis as a result of the phase transformation which is akin to the behavior observed with ferroelastic materials.  

While the results demonstrated in Fig.~\ref{fig4}d--e is only for a specified geometry, further tunability can be achieved by altering the geometric parameters. Here, we quantify the tunability by using the stiffness ratio $E_s^{(1)}/E_s^{(0)}$, where $E_s^{(1)}$ and $E_s^{(0)}$ are the initial, effective stiffness of the nonlinear stress-strain curve for the $100\%$ symmetric and $0\%$ symmetric, respectively. In Fig.~\ref{fig4}f, we plot the stiffness ratio as a function of $\ell_y/L_e$ for the specimens showing multistability. We observed that the deformed kirigami sheet with a coexistence configuration have a larger tunability in general. This tunability can be explained by the change of elastic energy stored in the kirigami lattice. When the generated lattice is in the symmetric mode, energy stored in the structure includes both out-of-plane bending energy ($U_b$) and elastic hinge energy ($U_h$). However, as we trigger the instability by supplying external energy to cross the energy barrier, the hinge energy would be dissipated. Absence of the hinge energy dominates the change of strain energy, furthermore, dominates the modulation of stiffness.

In summary, we showed that by controlling the slit spacing of the most simple kirigami pattern, linear parallel cuts, we can obtain a variety of multistable kirigami lattice structures which can be used to design architected materials with tunable properties. As an example, we show through simultaneously modifying the symmetry of the underlying lattice, a phase transformation occurs leading to a rapid and reversible modulation of the material stiffness. We construct a quantitative phase diagram with an analytic model to assist the rational selection of geometric parameters to design the multistable kirigami structure. Our results also indicate that the proposed design strategy is material independent. Although we focused on kirigami metamaterials at the meso-scale, this design strategy may be extended to different length scales\cite{Holmes2017}. We anticipate that embedding multistable kirigami into two-dimensional materials would establish a new platform to achieve {\em in situ} tunable mechanical, optical, and electrical properties.

\section{Acknowledgements}
YY and DPH gratefully acknowledge the financial support from NSF CMMI -- CAREER through Mechanics of Materials and Structures (\#1454153).



\end{document}